# *P-i-n* InGaN homojunctions (10-40% In) synthesized by plasma-assisted molecular beam epitaxy with extended photoresponse to 600 nm

S. Valdueza-Felip[1,2,3,*], A. Ajay[1,2], L. Redaelli[1,2], M. P. Chauvat[4], P. Ruterana[4],

T. Cremel[1,2], M. Jiménez-Rodríguez[1,2,3], K. Kheng[1,2], and E. Monroy[1,2]

[1] *University Grenoble-Alpes, 38000 Grenoble, France*

[2] *CEA, INAC-PHELIQS, 17 av. des Martyrs, 38000 Grenoble, France*

[3] *University of Alcalá (GRIFO), Mdr-Bcn Road, km 33.6, 28871 Alcalá de Henares, Spain*

[4] *CIMAP, CNRS-ENSICAEN-CEA-UCBN, 6 Blvd. Maréchal Juin, 14050 Caen, France*

We report the influence of the In mole fraction on the material and electrical characteristics of *p-i-n* $In_xGa_{1-x}N$ homojunctions ($x$ = 0.10–0.40) synthesized by plasma-assisted molecular-beam epitaxy on GaN-on-sapphire substrates. Junctions terminated with *p*-InGaN present improved carrier extraction efficiency in comparison with devices capped with *p*-GaN, due to the deleterious effect of polarization discontinuities on the device performance. We demonstrate that the presence of Mg does not perturb the In incorporation in InGaN, and it leads to a significant reduction of the stacking fault density. *p*-$In_{0.3}Ga_{0.7}N$ layers with a hole concentration of $3.2\times10^{18}$ cm$^{-3}$ are demonstrated. InGaN homojunction devices show a peak EQE = 14±2% in the blue-to-orange spectral region, and an extended cutoff to 600 nm.

---

[*] Email: sirona.valduezafelip@uah.es

# I. INTRODUCTION

As silicon solar cells approach their theoretical maximum efficiency, alternative materials for high-efficiency photovoltaics need to be investigated. A promising candidate is InGaN, whose direct bandgap tunable from 3.43 eV for GaN to 0.65 eV for InN covers nearly the entire solar spectrum. Moreover, only a few hundred nanometers of InGaN are required to absorb a large fraction of the incident light, thanks to its high absorption coefficient ($10^5$ cm$^{-1}$). This fact makes the realization of InGaN solar cells cost-effective, in spite of the presence of indium.

III-nitrides are interesting materials for high-efficiency multi-junction solar cells, in which several InGaN subcells with adjusted In content can be stacked to implement a full-spectrum-response device. Theoretical calculations point to a maximum efficiency of 62% for an all-InGaN device integrating 4 homojunctions [1]. Another multi-junction design consists of a *p-n* InGaN top cell connected in series to a *p-n* Si bottom cell through the *n*-In$_{0.45}$Ga$_{0.55}$N (1.8 eV)/*p*-Si (1.1 eV) interface, which shows the proper band alignment to form a low-resistance contact [2], or through an AlN, Si$_3$N$_4$ or SiO$_2$ interlayer [3]. This tandem device would benefit from the mature and low-cost Si technology, while its conversion efficiency could be boosted to 31% due to the nitride contribution at short wavelengths. Experimental evidence of the ohmic and rectifying behavior of this heterointerface has been reported for $In_xGa_{1-x}N$ ($x$ = 0.15−0.45) layers grown by molecular beam epitaxy (MBE) on *p*- and *n*-type Si(111), respectively [4].

However, growing an appropriate tunnel junction for $In_{0.5}Ga_{0.5}N$ on Si(111) is difficult due to the -12% lattice parameter mismatch and the 31% difference in thermal expansion coefficient. An alternative consist in stacking mechanically the two subcells, which might lead to an efficiency of 35% for a 600-nm-thick and 2.0-eV-bandgap InGaN top cell [5]. In this case photocurrent matching is no longer required since the two cells operate independently.

Regarding InGaN single junctions, various theoretical studies report on the effect of the In content, layer thickness, and defect density on the photovoltaic performance of InGaN homojunction and GaN/InGaN heterojunction solar cells [6,7,8]. By modelling the structure, a maximum efficiency of 17% could be achieved in an $In_{0.45}Ga_{0.55}N$-based device, dropping to 11% when capping it with a *p*-type GaN layer due to polarization issues in the *p-n* interface [8]. However, this theoretical promise has not been experimentally demonstrated yet. $In_xGa_{1-x}N$ homojunctions grown by metalorganic vapor phase epitaxy (MOVPE) achieve promising photovoltaic performance (peak external quantum efficiency EQE = 33%), but with an spectral cutoff limited to around 420 nm [9,10]. In contrast, MBE-grown $In_xGa_{1-x}N$ homojunctions with In concentrations reaching $x \leq 0.39$ have been fabricated, but there is no information on the EQE [11,12,13].

The successful fabrication of InGaN single-junction solar cells requires proper control of several key material issues such as (i) homogeneity of the In incorporation, (ii) reduction of structural defects such as threading dislocations and stacking faults, and (iii) proper Mg incorporation in the *p*-region for efficient acceptor concentration, and (iv) adequacy of the design to the presence of spontaneous and piezoelectric polarization.

In this work, we present the growth of Mg-doped InGaN films with 30% In displaying *p*-type conductivity with a hole concentration of $3.2 \times 10^{18}$ cm$^{-3}$. We demonstrate that the presence of Mg does not perturb the In incorporation in InGaN layers, and leads to a significant reduction of the stacking fault density. The material and electrical characteristics of In$_x$Ga$_{1-x}$N junctions with different In contents ($x$ = 0.10−0.40) and *p*-doped layers (GaN and InGaN) are investigated.

## II. EXPERIMENTAL

InGaN samples were grown by plasma-assisted MBE in a chamber equipped with standard effusion cells for Ga and In, Si and Mg for *n*- and *p*-type doping respectively, and a radio-frequency nitrogen plasma cell. Substrates consisted of 4-μm-thick GaN-on-sapphire templates. During the InGaN growth, the active nitrogen flux was fixed at $\Phi_N$ = 0.38 ML/s, and the InGaN growth temperature was in the 630–670 °C range. The growth was monitored *in situ* by reflection high-energy electron diffraction.

The surface morphology of the layers was evaluated by atomic force microscopy (AFM) using a Dimension 3100 system operated in the tapping mode. A structural analysis of the defects was performed via dark field transmission electron microscope (TEM) images recorded close to a <11-20> zone axis with g = 10-10 and g=0002. Optical transmission spectra were measured under normal incidence in the visible/near infrared spectral range (350−1700 nm) using an optical spectrum analyzer at room temperature.

The incorporation of Mg in the Mg-doped InGaN layers was quantified by secondary ion mass spectroscopy (SIMS), using $Cs^+$ and $O_2^+$ primary ion beams. Hall-effect measurements were performed at 300 K using the Van Der Pauw method. For this aim, Pd/Pt/Au (30 /40 /150 nm) ohmic contacts were placed at the edges of cross-shaped mesas etched down to the substrate.

The samples were processed into solar cell devices with mesa sizes of 0.5×0.5 and 1×1 mm$^2$ defined by $Cl_2$-based inductively coupled plasma etching. The *n*-InGaN contact surrounding the mesas consisted of e-beam evaporated Ti/Al/Ni/Au (30/70/20/100 nm). The *p*-(In)GaN contact consists of a semitransparent layer of Ni/Au (5/5 nm) layer annealed for 5 min in an oxygen-rich atmosphere at 500°C (96% transmittance measured in the visible spectral range), and a Ni/Au (30/100 nm) finger structure (finger width = 5 μm and pitch = 150 μm).

The spectral response of the devices was measured with a 1000 W halogen lamp coupled with an Omni Lambda 300 monochromator in the visible range (360–700 nm) calibrated with a reference Si photodetector. A Whitelase Supercontinuum SC-4x0 laser (repetition rate of 76 MHz, bandwidth of 10 nm tunable from 400 to 800 nm) was used to determine the external quantum efficiency (EQE) of the cells. Current density–voltage (*J–V*) measurements were recorded using an Agilent 4155C semiconductor parameter analyzer.

**III. RESULTS AND DISCUSSION**

InGaN layers and homojunctions were grown by plasma-assisted MBE in a chamber equipped with standard effusion cells for Ga, In, Si and Mg, and a radio-frequency nitrogen plasma cell. For the synthesis of InGaN, the fluxes were chosen as follows:

− the active nitrogen flux was fixed at $\phi_N = 0.38$ ML/s,

− the Ga flux at $\Phi_{Ga} = (1-x)\xi\Phi_N$, where *x* is the nominal In mole fraction, and $\xi = 0.74$ is the coefficient that accounts for the reduction of the growth rate associated to the InGaN decomposition, and

− the indium flux $\Phi_{In}$ was tuned to ensure a metal excess (*i.e.* $\Phi_{In} \gg x\,\Phi_N$).

The InGaN growth temperature was 640°C. With these growth conditions, a compositional gradient is expected during the first 350 nm of the growth due to the strain pulling effect, as experimentally demonstrated [14, 15].

A challenging issue for the fabrication of InGaN solar cells is the synthesis of high-quality In-rich *p*-type InGaN [16,17,18], due to the strong tendency of this material to be *n*-type doped through donor-type impurities and defects. Therefore, in a first stage, we have studied the incorporation of Mg in 830-nm-thick $In_{0.3}Ga_{0.7}N$ layers grown on insulating GaN:Fe-on-sapphire substrates by varying the Mg temperature in the 250-300°C range as indicated in Table I (samples D1-D3). An identical sample non-intentionally doped (*n.i.d*) was also grown to serve as a reference (sample D0). SIMS measurements were performed in a reference sample containing alternating layers of *n.i.d* $In_{0.3}Ga_{0.7}N$ and $In_{0.3}Ga_{0.7}N$:Mg layers grown at various Mg cell temperatures. The results are summarized in Table I.

The values of carrier concentration measured by Hall-effect at 300 K using the Van Der Pauw method are summarized in Table I. The reference sample D0 presents a residual electron concentration of $n = 4.4 \times 10^{17}$ cm$^{-3}$. A hole concentration $p = 3.2 \times 10^{18}$ cm$^{-3}$ is obtained at low Mg fluxes ($T_{Mg} = 250$ °C, corresponding to a Mg incorporation $[Mg] = 5 \times 10^{18}$ cm$^{-3}$). Increasing the Mg flux to $T_{Mg} = 275$°C and above results in self-compensation (D2) or even *n*-type conductivity (D3), indicating that the excess of Mg doping leads to the formation of donor-type defects, as previously reported for Mg-doped GaN [17].

Dark field TEM images of samples D0 and D1 recorded close to a <11-20> zone axis with $g = 10\text{-}10$ in Fig. 1 reveal that the incorporation of Mg leads to a significant reduction of the density of stacking faults [19] and dislocations [20].

With this information on the Mg doping, we grew a series of *p-i-n* $In_xGa_{1-x}N$ homojunctions with nominal thickness of $p = 80$ nm, $i = 50$ nm and $n = 500$ nm and nominal In mole fractions of $x = 0.10, 0.20, 0.30$ and $0.40$, as described in Table II. The thickness of the *n* layer is chosen so as to attain strain relaxation (to prevent strain pulling effects in the active area [14]), and the *p* layer corresponds approximately to the minority carrier diffusion length in Mg-doped GaN [21]. The substrates consisted of commercial 4-µm-thick Si-doped GaN deposited on sapphire by MOVPE. The growth conditions were the same as in the previous series, except for the InGaN growth temperature, which was set to 667ºC, 648 ºC, 636ºC and 629ºC for samples S1-S4, respectively (see Table II) [14]. The *n*-type layer was doped with Si at a concentration of $5 \times 10^{18}$ cm$^{-3}$. The *i* layer was non-intentionally doped. Regarding the *p*-type layer, the Mg cell temperature was tuned as a function of the growth temperature (*i.e.* the targeted In content) to compensate Mg segregation/desorption and achieve *p*-type conductivity. The Mg, Ga and In depth profiles measured by SIMS, shown in Figs. 2(a) and (b) for samples S1 and S4, respectively, indicate that the Mg atoms are uniformly incorporated throughout the *p*-layer, and the In incorporation does not decrease with the presence of Mg atoms, in contrast with previous reports [17]. Regarding the In profile, for all In contents under study, we observe a gradual increase of the In mole fraction with the growing thickness in the first ~300 nm. This compositional gradient is attributed to the strain relaxation, as recently reported for In$_{0.3}$Ga$_{0.7}$N layers grown by plasma-assisted MBE [14]. This effect is reduced for lower In contents [sample S1, Fig. 2(a)], with a smaller lattice mismatch between the InGaN structure and the GaN substrate. The In mole fraction was estimated from SIMS measurements using reference samples of GaN doped with In, with the results summarized in Table II. Note that with this measurement procedure, the quantitative evaluation of the In content obtained by SIMS is only accurate for low In compositions.

Fig. 3 displays the AFM images of the topmost InGaN:Mg in the homojunctions (samples S1 to S4) showing smooth surfaces with no pits, and a root-mean-squared (rms) roughness as low as 1.3–2.0 nm. These values are similar to previous results obtained on thick *n.i.d* InGaN layers [14].

Fig. 4(a) shows the transmission spectra for the homojunctions with various In concentrations, where the absorption cutoff redshifts accordingly with the increasing In content. The bandgap energy ($E_g$) was estimated from the Tauc's plots in Fig. 4(b), which represents the energetic dependence of $(\alpha E)^2$, where $E$ is the impinging photon energy and $\alpha$ the absorption coefficient. The values of $E_g$, summarized in Table II, redshift from 2.75 eV (450 nm) to 2.25 eV (550 nm) for samples S1 to S4, respectively.

The tail of density of states in the InGaN material due to the alloy inhomogeneity is associated to an absorption edge broadening $\Delta E$. To quantify this value, the variation of the absorption coefficient as a function of the photon energy is fitted to a sigmoidal function:

$$\alpha = \frac{\gamma}{1 + \frac{E_0 - E}{\Delta E}}$$

where γ is a fitting parameter, $\Delta E$ is the absorption edge broadening (equivalent to the Urbach tail energy), and $E_0$ is a fitting parameter which provides an upper estimation of the average bandgap energy. The value of $\Delta E$ increases with the In content from 53±1 meV to 66±1 meV.

A sketch of the devices together with a top-view optical microscopy image is shown in Fig. 5(a) and (b), respectively. Further details about the fabrication process are reported in [22].

The EQE of the solar cells was calculated in the 480–590 nm wavelength range as

$$EQE = \frac{J_{op}}{P_{op}} \frac{hc}{q\lambda},$$

where $J_{op}$ is the photocurrent density, $P_{op}$ the optical power density impinging the device, $q$ the electron charge, $h$ is Planck's constant, $c$ is the speed of light, and $\lambda$ the wavelength of the incident light.

Fig. 4(c) presents the spectral response of various devices with different In contents. The EQE cutoff wavelength ($\lambda_{cutoff}$) is estimated from the linear interpolation of the spectrum edge with the x-axis in linear scale [see Fig. 7(a)]. They present a flat photoresponse in the blue-to-green spectral region with a sharp cutoff which redshifts from $\lambda_{cutoff}$ = 465 nm (S1) to $\lambda_{cutoff}$ = 600 nm (S4), as expected from the estimated bandgaps of the absorbing InGaN material. The spectral shift of the devices within one sample scales with the In content from 5 in S1 to 25 nm in S4, as illustrated in Fig. 5(c). This variation is attributed to the In content inhomogeneity on each wafer, which increases for higher In contents accordingly. Peak responsivity values are 19 mA/W (S1), 66 mA/W (S2), 58 mA/W (S3) and 50 mA/W (S4), measured at 445, 480, 510 and 525 nm, respectively. Peak EQE values are summarized in Table II. These values are below the state-of-the art of $In_{0.15}Ga_{0.75}N$ homojunctions grown by MOVPE (EQE = 33%), but in this case the spectral cutoff is limited to ~420 nm [10]. On the contrary, our EQE results are higher than those recently obtained on $GaN/In_{0.20}Ga_{0.80}N$ heterojunctions grown by MBE with a spectral cutoff of ~515 nm (EQE ~ 10%) [23].

The difficulties to achieve high *p*-type conductivity in high-In-content InGaN layers have lead many groups to replace the *p*-InGaN by a *p*-GaN layer [10, 24]. In order to evaluate the difference between both strategies, two $In_{0.30}Ga_{0.70}N$ structures with nominal thicknesses of $p = 120$ nm, $i = 120$ nm and $n = 960$ nm, and different *p*-type top layers, namely GaN:Mg (junction J1), and $In_{0.30}Ga_{0.70}N$:Mg (junction J2) were grown. In J1, the substrate and the Mg cell temperatures were respectively raised to $T_s = 720°C$ and $T_{Mg} = 325°C$ for the growth of the GaN:Mg, compared to J2. For these growth conditions, the hole concentration of the GaN layer is $p = 5 \times 10^{17}$ cm$^{-3}$ ([Mg] = $1.3 \times 10^{19}$ cm$^{-3}$) [25]. As expected, Mg-doped InGaN has a higher hole concentration than Mg-doped GaN since the Mg activation energy decreases with the bandgap energy and thus the In mole fraction [17]. Fig. 6 shows cross-section dark field TEM images of J1 and J2 viewed close to a <11-20> zone axis with g=0002. The strong lattice mismatch between the InGaN and the top *p*-GaN in J1 results in the appearance of defects and a strain-related contrast at the heterointerface.

The spectral dependence of the EQE for both junctions is presented in Fig. 7(a). The GaN-terminated structure (sample J1) presents a strong photocurrent cutoff at $\lambda_{cutoff}$ = 375 nm, with a peak responsivity of 58.3 mA/W (EQE = 18.1%) at 350 nm. This spectral response points to a photocurrent generated only in the *p*-GaN layer with no contribution from the InGaN material. This can be explained by deficient hole collection, which is justified by the band profile of the samples in Fig. 7(b) (calculated at zero bias with the nextnano$^3$ software [26]), where a blocking barrier for hole transport into GaN is generated at the GaN/InGaN interface due to the difference in polarization, as illustrated by a dotted circle in Fig. 7(b). This potential barrier could be reduced by inserting a lightly doped thin layer [27].

On the contrary, the In$_{0.30}$Ga$_{0.70}$N homojunction (J2) shows a flat photoresponse in the blue-to-green spectral region with a sharp cutoff redshifted to $\lambda_{cutoff}$ = 535 nm, as expected from the band diagram in Fig. 7(b). The observed behavior is in good agreement with the simulations performed by Fabien *et al.* for single InGaN/GaN hetero- and InGaN homojunction solar cells [5]. The peak responsivity is 48.6 mA/W (EQE = 12.5%) at 485 nm for the all-InGaN junction (J2), and it decreases to 24.5 mA/W (EQE = 8.7%) at 350 nm due to carrier recombination within the *p*-InGaN layer. The EQE drop at short wavelength (300–350 nm range) for both junctions is likely related to carrier recombination at the surface.

The *J-V* curve in the dark of J2 in the inset of Fig. 7(a) shows rectifying behavior with an average series resistance $R_S$ = 0.27±0.05 $\Omega$cm$^2$ and shunt resistance $R_{Sh}$ = 0.75±0.25 M$\Omega$cm$^2$.

A photovoltaic response is observed under illumination with a halogen lamp (average optical power = 19 mW/cm$^2$).

## III. CONCLUSIONS

In$_x$Ga$_{1-x}$N homojunctions ($x$=0.10–0.40) were fabricated by plasma-assisted molecular-beam epitaxy and compared to similar *p*-GaN terminated structures. The presence of Mg does not perturb the In incorporation in InGaN layers and leads to a strong reduction of the stacking fault density. *p*-In$_{0.30}$Ga$_{0.70}$N layers with a hole concentration of 3.2×10$^{18}$ cm$^{-3}$ are demonstrated. *P*-GaN terminated devices show reduced extraction efficiency due to polarization-related potential barriers that are avoided using *p*-InGaN top layers. All-InGaN devices present a broad spectral response covering the blue-to-orange region with peak values of 14±2%, and a sharp cutoff that redshifts with increasing In content from 465 nm to 600 nm.

**Acknowledgements**

Financial support was provided by the Marie Curie IEF grant "SolarIn" (#331745), and by the French National Research Agency via the GANEX program (ANR-11-LABX-0014). Devices were fabricated in the Plateforme Technologique Amont (PTA) of the CEA in Grenoble, France.

**Tables**

**Table I:** Summary of the In$_{0.3}$Ga$_{0.7}$N samples doped with Mg including the Mg cell temperature (T$_{Mg}$), the Mg concentration [Mg] measured by SIMS in a reference sample grown under the same conditions, and the carrier concentration extracted from Hall effect measurements.

| In$_{0.3}$Ga$_{0.7}$N samples | T$_{Mg}$ (ºC) | Doping type | [Mg] (cm$^{-3}$) | Carrier concentration (cm$^{-3}$) |
|---|---|---|---|---|
| D0 | 0 | *n* | - | 4.4×10$^{17}$ |
| D1 | 250 | *p* | 5×10$^{18}$ | 3.2×10$^{18}$ |
| D2 | 275 | *i* | 7×10$^{18}$ | - |
| D3 | 300 | *n* | 1.5×10$^{19}$ | 2.9×10$^{19}$ |

**Table II:** Description of the *p-i-n* In$_x$Ga$_{1-x}$N homojunctions comprising nominal In mole fraction (*x*), growth temperature (T$_{substrate}$), Mg temperature (T$_{Mg}$) and rms surface roughness. Mg concentration in the InGaN *p*-type layer and In mole fraction ($x_{min}$ minimum, $\bar{x}$ average, and $x_{max}$ maximum) measured by SIMS. Bandgap energy ($E_g$) extracted from Tauc's plot. Fitting parameters of the absorption coefficient using a sigmoidal approximation ($E_0$ and $\Delta E$). Peak EQE and EQE cutoff ($\lambda_{cutoff}$) extracted from photocurrent experiments are also included.

| Homo junction | Nominal In mole fraction $x$ | $T_{substrate}$ (ºC) | $T_{Mg}$ (ºC) | Rms roughness (nm) | [Mg] (cm$^{-3}$) | In mole fraction | | | $E_g$ (eV) | Sigmoidal approximation | | Peak EQE (%) | $\lambda_{cutoff}$ (nm) |
|---|---|---|---|---|---|---|---|---|---|---|---|---|---|
| | | | | | | $x_{min}$ | $\bar{x}$ | $x_{max}$ | | $E_0$ (eV) | $\Delta E$ (meV) | | |
| S1 | 0.10 | 667 | 300 | 1.3 | $1.5\times10^{19}$ | 0.11 | 0.13 | 0.15 | 2.75 | 2.85 | 54 | 5.5 | 465 |
| S2 | 0.20 | 648 | 275 | 1.8 | $7.2\times10^{18}$ | 0.15 | 0.20 | 0.24 | 2.63 | 2.71 | 52 | 17 | 510 |
| S3 | 0.30 | 636 | 250 | 1.9 | $5.5\times10^{18}$ | - | - | - | 2.34 | 2.41 | 62 | 14 | 560 |
| S4 | 0.40 | 629 | 250 | 2.0 | $6.8\times10^{18}$ | - | - | - | 2.25 | 2.30 | 66 | 12 | 600 |

**Figure Captions**

Figure 1: TEM images of the (a) *n.i.d* and (b) Mg-doped ($T_{Mg}$ = 250°C) *p*-type $In_{0.3}Ga_{0.7}N$ samples grown on GaN:Fe-on-sapphire substrates (D0 and D1, described in Table I).

Figure 2: Mg concentration and Ga-In profiles of the $In_xGa_{1-x}N$ homojunctions (a) S1 ($x = 0.10$) and (b) S4 ($x = 0.40$) measured by SIMS.

Figure 3: AFM images (2×2 µm$^2$ size) of the InGaN homojunctions (S1-S4 in Table II).

Figure 4: Room-temperature transmittance (a) and Tauc's plot (b) of the InGaN homojunctions (S1-S4 in Table II).

Figure 5: (a) Schematic description of the *p-i-n* InGaN solar cell structure. (b) Top-view optical microscopy image of a device with 0.5×0.5-mm$^2$ mesa. (c) External quantum efficiency spectra of the InGaN homojunctions (S1-S4 described in Table II). The different lines correspond to different solar cells within one sample (15×15 mm$^2$). There is no clear correlation between the location of the cell and its cutoff wavelength.

Figure 6: TEM images of the GaN/InGaN heterojunction and the InGaN homojunction with 30% of In with a total thickness of 1210 nm and 1300 nm for J1 and J2, respectively.

Figure 7: (a) External quantum efficiency, and (b) band energy profile of the GaN/InGaN heterojunction and the InGaN homojunction with 30% of In (J1 and J2, respectively). In (b), the hole potential barrier generated by the polarization difference at the heterointerface is outlined with a dashed circle. Inset of Fig. 7(a): Current density vs voltage curves of J2 in the dark and under halogen illumination (average power density = 19 mW/cm$^2$).

**Figure 1**

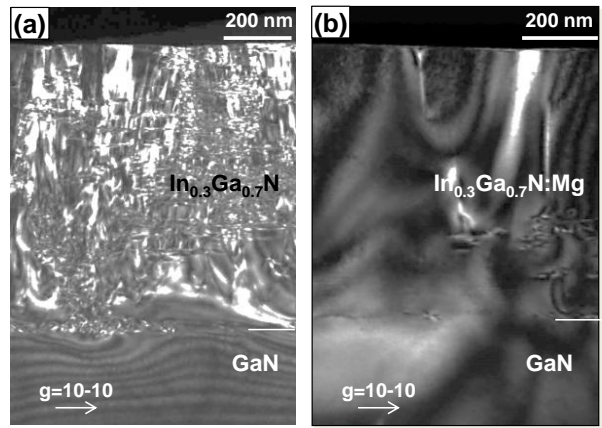

**Figure 2**

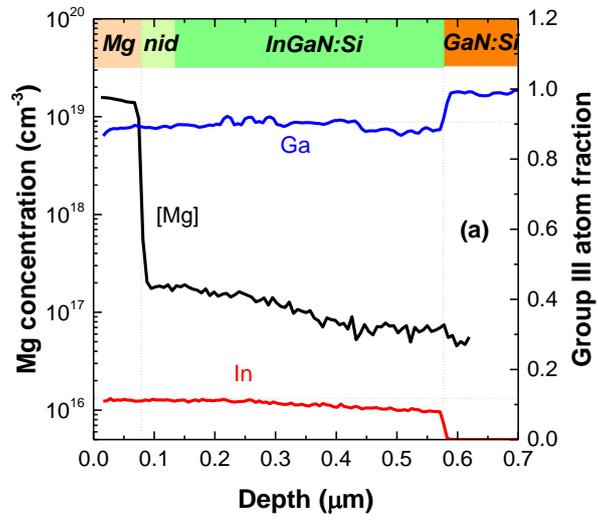

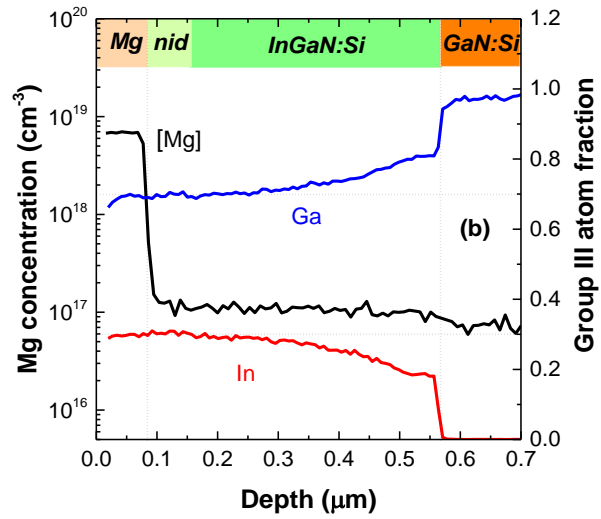

**Figure 3**

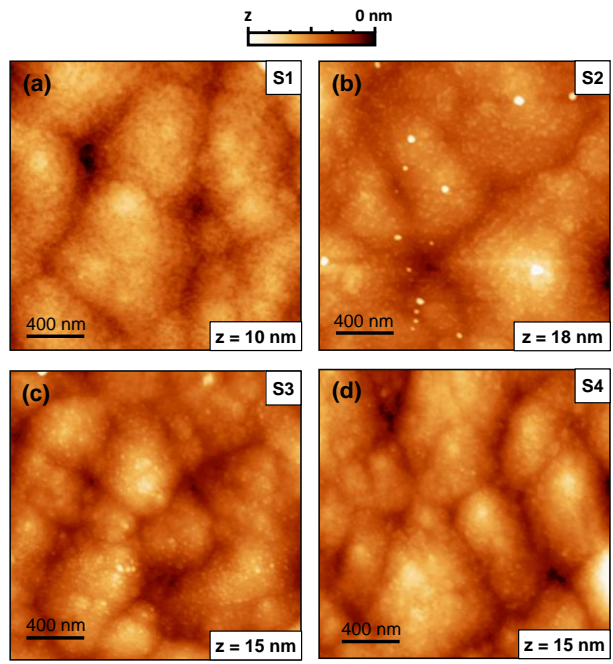

**Figure 4**

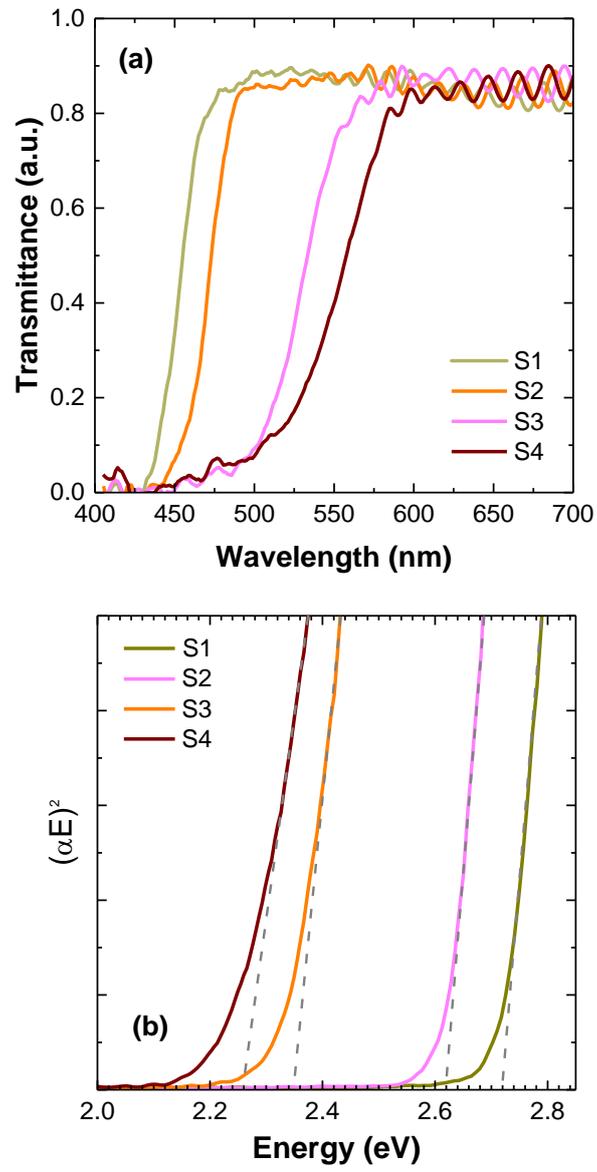

**Figure 5**

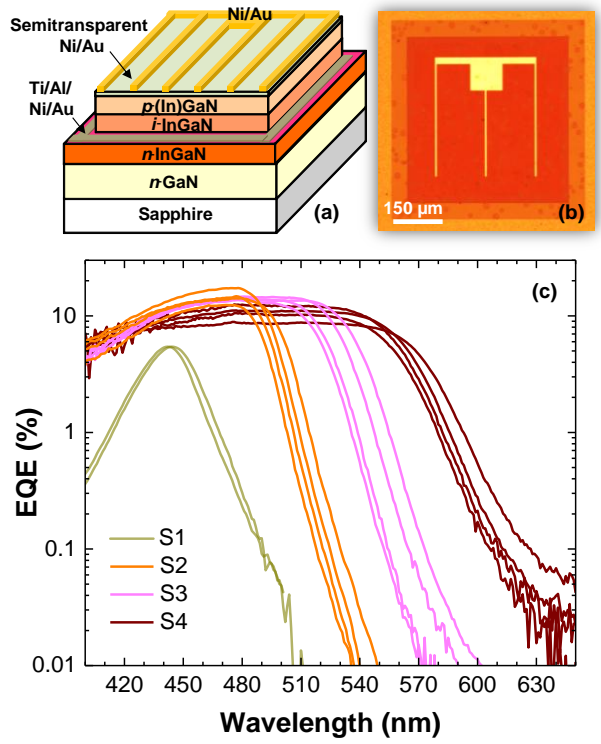

**Figure 6**

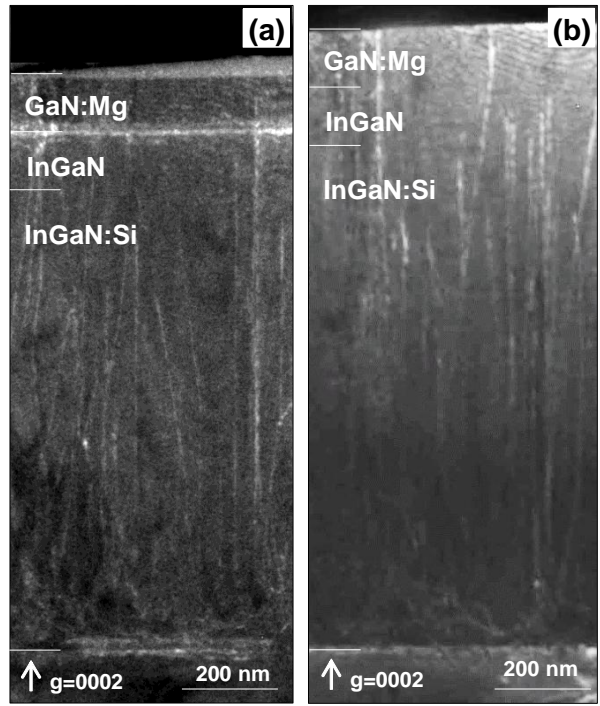

**Figure 7**

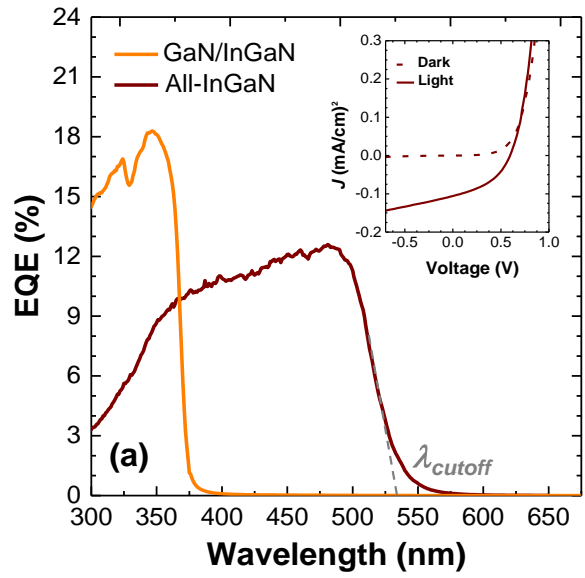

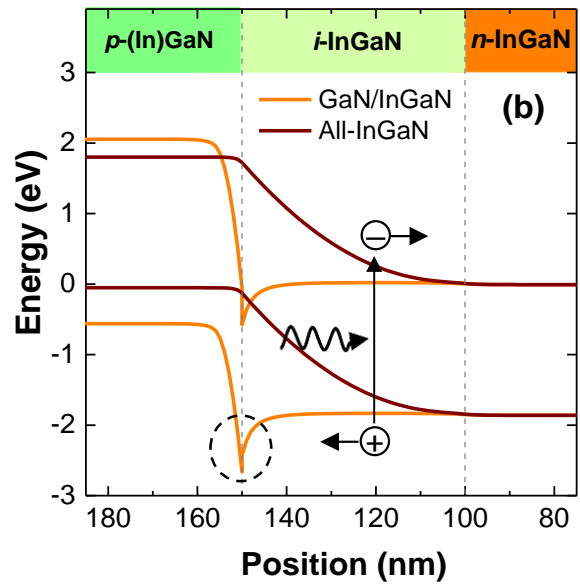